\documentstyle[aps,prbbib,floats]{revtex} \hoffset=-0.6cm
\begin{document}
\wideabs{
\title{Antiferromagnetic exchange and spin-fluctuation pairing
\\ in cuprate superconductors}
\author{N. M. Plakida
}
\address{Joint Institute for Nuclear Research,
 141980 Dubna,  Russia}
\maketitle
\begin{abstract}
A microscopic theory of superconductivity  is formulated within an
effective $p$-$d$ Hubbard model for a CuO$_2$ plane. By applying
the Mori-type projection technique, the Dyson equation is derived
for the Green functions in terms of  Hubbard operators. The
antiferromagnetic exchange caused by interband hopping results in
pairing of all carries in the conduction subband and high $T_c$
proportional to the Fermi energy.  Kinematic interaction in
intraband hopping is responsible for the conventional
spin-fluctuation pairing. Numerical solution of the gap equation
proves  the $d$-wave gap symmetry and defines $T_c$ doping
dependence. Oxygen isotope shift and pressure dependence of $T_c$
are also discussed.
\end{abstract}
 }

\section{INTRODUCTION}

A unique property of cuprates is the huge antiferromagnetic (AFM)
superexchange interaction $J \simeq 1500$~K which causes a
long-range AFM order in the undoped regime, while in the metallic
state results in strong AFM dynamical spin fluctuations. These are
responsible for anomalous normal state properties of cuprates and
for superconducting pairing as  originally proposed by
P.W.~Anderson~\cite{Anderson87}. In  a number of studies of the
reduced  one-band $t$-$J$ model (see,
e.g.~\cite{Plakida99,Plakida01} and references therein) it was
shown that the  AFM exchange interaction results in $d$-wave
pairing with  high $T_c$.  However, to asses the origin of the AFM
pairing mechanism one has to consider the original two-band
$p$-$d$ model for CuO$_2$ plane without reducing it  to an
effective $t$-$J$ model with  instantaneous exchange interaction
in one subband by projecting out the intersubband hopping. In
fact, this transformation is used in derivation of the BCS model
with an instantaneous electron interaction by projecting out
electron-phonon coupling in the original electron-ion model.
\par
In this paper  a microscopic theory of superconductivity within
the effective $p$-$d$ Hubbard model is presented. By applying a
projection technique for the  matrix Green function in terms of
Hubbard operators the Dyson equation is derived~\cite{Plakida03}.
It is found that  the mean-field solution results in   $d$-wave
superconducting pairing mediated by the exchange interaction due
to interband hopping, similar to  the $t$-$J$ model. The
self-energy caused by the kinematic interaction in the intraband
hopping is calculated in the non-crossing approximation. It
defines the conventional spin-fluctuation pairing.  Numerical
solution of the gap equation gives the doping dependence of
superconducting $T_c$ and wave-vector dependence of the 
gap   which are in agreement with experiments~\cite{Plakida03}.
$T_c$ increase under pressure   and  oxygen isotope effect in
cuprates are also discussed~\cite{Plakida01}.

\section{DYSON EQUATION}

Starting from the original two-band $p$-$d$ model for  CuO$_2$
plane we can reduce it to an effective two-band Hubbard model  by
applying the cell-cluster perturbation theory (see,
e.g.~\cite{Plakida95}):
\begin{eqnarray}
H &= &  E_1\sum_{i,\sigma}X_{i}^{\sigma \sigma} +
E_2\sum_{i}X_{i}^{22} + \sum_{i\neq
j,\sigma}\{t_{ij}^{11}X_{i}^{\sigma 0}X_{j}^{0\sigma}
\nonumber\\
& + &  t_{ij}^{22}X_{i}^{2 \sigma}X_{j}^{\sigma 2} +2\sigma
t_{ij}^{12}(X_{i}^{2\bar\sigma}X_{j}^{0 \sigma} + {\rm H.c.})\},
\label{eq:H}
\end{eqnarray}
where $X_{i}^{nm} = |in\rangle\langle im|$ are  Hubbard operators
for  four states $n,m=|0\rangle ,\,|\sigma\rangle ,\, |2\rangle
=|\uparrow \downarrow \rangle $, $\sigma=\pm 1/2 $,
$\bar\sigma=-\sigma$.  $E_1=\epsilon_d-\mu$ and
$\,E_2=2E_1+\Delta\,$  are  energy levels of the lower Hubbard
subband  (one-hole Cu $d$-like states) and the upper Hubbard
subband (two-hole $p$-$d$ singlet states), respectively.$\;$
 $\mu$ is the chemical potential,
$\Delta=\epsilon_p-\epsilon_d \sim 3$~eV is the charge transfer
energy and $t \sim \Delta/2 $ is the $p$-$d$ hybridization
parameter. The hopping integrals $ \mid t^{\alpha\beta}_{i j}\mid
\ll \Delta \,$,  e.g. $\, t^{22}_{i, i\pm a} = t_{nn} \simeq 0.14
t \sim 0.2$~eV, which means that the Hubbard model~(\ref{eq:H})
corresponds to a strong correlation limit. We treat it within the
Hubbard operator technique to preserve rigorously  a constraint of
no double occupancy of any quantum state $|in\rangle$ which
follows from the completeness relation: $ X_{i}^{00} +
X_{i}^{\sigma\sigma} + X_{i}^{\bar\sigma\bar\sigma}
 + X_{i}^{22} = 1$.
\par
To discuss the superconducting pairing we introduce the
four-component Nambu operators $\, \hat X_{i\sigma}$ and $\, \hat
X_{i\sigma}^{\dagger}$ and  define the $4\times 4$ matrix Green
function (GF)~\cite{Zubarev60}:
\begin{eqnarray}
\tilde G_{ij\sigma}(t-t')& =& \langle\langle \hat X_{i\sigma}(t)\!
\mid \!  \hat X_{j\sigma}^{\dagger}(t')\rangle\rangle ,
\nonumber\\
 \tilde G_{ij\sigma}(\omega)& = &{\hat G_{ij\sigma}(\omega)  \quad \quad
 \hat F_{ij\sigma}(\omega) \choose
 \hat F_{ij\sigma}^{\dagger}(\omega) \quad
   -\hat{G}_{ji\bar\sigma}(-\omega)} ,
 \label{eq:GF}
\end{eqnarray}
where $\, \hat X_{i\sigma}^{\dagger}=(X_{i}^{2\sigma}\,\,
X_{i}^{\bar\sigma 0}\,\, X_{i}^{\bar\sigma 2}\,\, X_{i}^{0\sigma})
\,$.  $\hat G_{ij\sigma}$ and  $\hat F_{ij\sigma}$ are the
two-subband normal and anomalous  matrix components, respectively.
By applying the projection technique for the equation of motion
method for GF~(\ref{eq:GF}) we derive the Dyson equation in the
$({\bf q},\omega)$-representation \cite{Plakida03}:
\begin{eqnarray}
\left( \tilde G_\sigma({\bf q},\omega) \right)^{-1}& = &\left(
\tilde G_{\sigma}^{0}({\bf q}, \omega) \right)^{-1} -
\tilde\Sigma_{\sigma}({\bf q}, \omega),
\nonumber\\
\tilde G^0_\sigma ({\bf q},\omega)& = & \Bigl(\omega \tilde
\tau_{0} - \tilde E_{\sigma}({\bf q}) \Bigr)^{-1}\tilde \chi ,
  \label{eq:Dyson}
\end{eqnarray}
where $\tilde\chi=\langle\{\hat X_{i\sigma},\hat
X_{i\sigma}^{\dagger}\}\rangle \,$. The zero-order GF within the
generalized mean field approximation (MFA) is defined by the
frequency matrix which in the site representation  reads
\begin{equation}
\tilde E_{ij\sigma} = \tilde {\cal A}_{ij\sigma} \tilde\chi^{- 1},
\quad \tilde {\cal A}_{ij\sigma} = \langle\{[\hat
X_{i\sigma},H],\hat X_{j\sigma}^{\dagger}\}\rangle.
 \label{eq:Aij}
\end{equation}
 The self-energy operator in the Dyson equation
(\ref{eq:Dyson}) in the projection technique method is defined by
a {\it proper } part (having no single zero-order GF) of the
many-particle GF in the form
\begin{equation}
\tilde \Sigma_{\sigma}({\bf q}, \omega) = {\tilde\chi}^{-1}
\langle\!\langle {\hat Z}_{\sigma}^{(ir)} \!\mid\!  {\hat
Z}_{\sigma}^{(ir)\dagger} \rangle\!\rangle^{(prop)}_{{\bf q},
\omega}\;{\tilde\chi}^{-1}.
 \label{eq:self-enir-qo}
\end{equation}
Here the {\it irreducible} $\hat Z $-operator is given by the
equation: $ {\hat Z}_{\sigma}^{(ir)} = [\hat X_{i\sigma},H] -
\sum_{l}\tilde E_{il\sigma} \hat X_{l\sigma}$ which follows from
the orthogonality condition:  $ \langle\{{\hat
Z}_{\sigma}^{(ir)},{\hat X}_{j\sigma}^{\dagger}\}\rangle = 0$. The
equations~(\ref{eq:Dyson})-(\ref{eq:self-enir-qo}) provide an
exact representation for the
 GF~(\ref{eq:GF}). Its calculation, however, requires the
use of some approximation for the self-energy
matrix~(\ref{eq:self-enir-qo}) which describes  finite lifetime
effects (inelastic scattering of electrons on spin and charge
fluctuations).

\section{MEAN-FIELD SOLUTION}

In the MFA the electronic spectrum and superconducting pairing are
described by the zero-order GF in Eq.~(\ref{eq:Dyson}) where the
frequency matrix (\ref{eq:Aij}) reads:
\begin{equation}
\tilde {\cal A}_{ij\sigma} = { \hat{\omega}_{ij\sigma} \quad \quad
\hat{\Delta}_{ij\sigma} \choose \hat{\Delta}_{ji\sigma}^{*} \quad
-\hat{\omega}_{ji\bar\sigma} }. \label{aij}
\end{equation}
Here $\hat{\omega}_{ij\sigma}$ and $\hat{\Delta}_{ij\sigma}$ are
$\, 2\times 2 \, $ matrices for the normal and anomalous
components, respectively. The normal component defines
quasiparticle spectra of the model in the normal state which have
been  studied in detail in~\cite{Plakida95}. The anomalous
component defines the gap functions, e.g., for the singlet subband
we have:
\begin{equation}
  \Delta\sb{ij\sigma}\sp{22}  =  - 2 \sigma t_{ij}^{12}
   \langle X\sb{i}\sp{02} N\sb{j} \rangle,\;(i \neq j)
 \label{l22}
\end{equation}
where  $\, N\sb{i}  = \sum\sb{\sigma} X\sb{i}\sp{\sigma \sigma} +
2  X\sb{i}\sp{22}$  is the number operator.  In terms of the Fermi
operator: $ c_{i\sigma} = X_{i}^{0\sigma} + 2\sigma
X_{i}^{\bar\sigma 2}$ we can write the anomalous average as
$\,\bigl<c\sb{i\downarrow}c\sb{i\uparrow}  N\sb{j} \bigr> = \bigl<
X\sb{i}\sp{0\downarrow} X\sb{i}\sp{\downarrow 2}  N\sb{j}\bigr> =
  \bigl< X\sb{i}\sp{02} N\sb{j}\bigr> \, $ since other products
of  Hubbard operators vanish according to  multiplication rules
for them:
 $\, X_{i}^{\alpha\gamma}X_{i}^{\lambda\beta}
  =  \delta_{\gamma,\lambda} \, X_{i}^{\alpha\beta}\,$.
Therefore, the anomalous correlation functions describe the
pairing at one lattice site but in different subbands.
 \par
 The same anomalous correlation functions
were obtained in MFA for the original Hubbard model in
Refs.~\cite{Beenen95,Avella97,Stanescu00}. To calculate them
in~\cite{Beenen95,Stanescu00} the Roth procedure was applied based
on  decoupling of  operators on the same lattice site in the
time-dependent correlation function:
 $\langle c_{i\downarrow}(t)|c_{i\uparrow}(t')N_j (t')\rangle \,$.
However, the decoupling of Hubbard operators on one lattice site
is not unique and therefore unreliable. The use of kinematic
restrictions for the Hubbard operators~\cite{Avella97} instead of
decoupling also did not produce a unique  solution.
\par
We calculated  the correlation function $\bigl< X_i^{02} N_{j}
\bigr>$ without {\it any decoupling} from the equation of motion
of the corresponding commutator GF
 $ L_{ij}(t-t') = \langle\langle X_{i}^{02} (t)
  \mid N_j (t') \rangle \rangle \, $.
By neglecting exponentially small terms of the order of $\, \exp(-
\Delta/T)\,$ we obtain  for the correlation function in the
singlet subband ($i \ne j $)~\cite{Plakida03}:
\begin{eqnarray}
  \langle X_{i}^{02} N_j \rangle & = &
  - \, ({1}/{\Delta }) \, \sum_{m\neq i,\sigma} \, 2\sigma t^{12}_{im}
  \langle X_{i}^{\sigma 2} X_{m}^{\bar\sigma 2}  N_j \rangle
\nonumber \\
 & \simeq &
   - \,  ({4 t^{12}_{ij} }/{\Delta }) 2 \sigma \,
  \langle X_{i}^{\sigma2}X_{j}^{\bar\sigma2}\rangle .
 \label{eq:x2d}
\end{eqnarray}
The approximate value  is derived in the two-site approximation
usually applied for the $t$-$J$ model:  $\, X_{m=j}^{\bar\sigma 2}
N_j = 2 X_{j}^{\bar\sigma 2}\,$. Therefore, the  gap equation
(\ref{l22}) in the case of hole doping reads
\begin{eqnarray}
\Delta^{22}_{ij\sigma}  = -  2 \sigma \, t^{12}_{ij}
 \langle X_{i}^{02} N_j \rangle =
  J_{ij}  \langle X_{i}^{\sigma2}X_{j}^{\bar\sigma2}\rangle ,
 \label{eq:x2e}
\end{eqnarray}
where $\, J_{ij} = {4\, (t^{12}_{ij})^2}/{\Delta}$. It is  the
conventional  exchange interaction  pairing  in the $t$-$J$ model.
 For electron doping analogous  calculations for the anomalous
correlation function $\, \langle (2 - N\sb{j}) X\sb{i}\sp{02}
\rangle \,$ gives for the gap function $\, \Delta_{ij\sigma}^{11}=
J_{ij}\, \langle X_{i}^{0\bar\sigma} X_{j}^{0\sigma} \rangle $.
\par
We  conclude that the anomalous contributions to the zero-order
GF, Eq.~(\ref{eq:Dyson}), are described by conventional pairs of
quasi-particles in one subband and there are no new composite
operator excitations, the "cexons",  proposed
in~\cite{Stanescu00}.

\section{SELF-ENERGY}

The  self-energy matrix~(\ref{eq:self-enir-qo}) can be written in
the form
\begin{equation}
\tilde \Sigma_{ij\sigma}(\omega) = \tilde\chi^{-1} {\hat
M_{ij\sigma}(\omega) \quad  \quad \hat\Phi_{ij\sigma}(\omega)
\choose \hat\Phi_{ij\sigma}^{\dagger} (\omega)\quad
-\hat{M}_{ij\bar\sigma}(-\omega)} \tilde \chi^{-1} \, ,
\label{eq:sigma-ro}
\end{equation}
where the $2\times 2$ matrices $\hat M$ and $\hat\Phi$ denote the
normal and anomalous contributions to the self-energy,
respectively. We calculate them   in the non-crossing, or the
self-consistent Born approximation (SCBA).  In SCBA, Fermi-like
and Bose-like excitations in the many-particle GF
in~(\ref{eq:sigma-ro}) are assumed to  propagate independently
which is given by  the  decoupling of the corresponding
time-dependent correlation functions as follows:
\begin{eqnarray}
 && \langle B_{1'}(t) X_{1}(t) B_{2'}(t') X_{2}(t')
 \rangle|_{(1 \neq 1', 2 \neq 2')}
\nonumber \\
&\simeq &\langle X_{1}(t) X_{2}(t') \rangle \langle B_{1'}(t)
B_{2'}(t') \rangle. \label{scba}
\end{eqnarray}
Using the spectral representation for the correlation functions
 we get a closed system of equations for the GF~(\ref{eq:GF})
and the self-energy~(\ref{eq:sigma-ro}). For instance, the
anomalous part of the self-energy for the singlet subband reads:
\begin{eqnarray}
\Phi_{\sigma}^{22}({\bf q},\omega) = \frac{1}{2\pi N}\sum_{{\bf
k}} \int\!\!\!\int_{-\infty}^{+\infty}
 \frac{{\rm d}\omega_1 {\rm d}\omega_{2}}
{\omega-\omega_1-\omega_2}   \nonumber \\
 \left( \tanh{{\omega_1}/{2T}}
  + \coth {{\omega_2}/{2T}}\right)\;
   \chi_{s}^{\prime\prime}({\bf q-k}, \omega_2)\quad
 \label{eq:phi-qo}\\
\{ -{\rm Im}[t_{22}^2({\bf k})
 F_{\sigma}^{22}({\bf k},\omega_1) - t_{11}^2({\bf k})
   F_{\sigma}^{11}({\bf k},\omega_1)]\} .
\nonumber
\end{eqnarray}
Here the spin-fluctuation  pairing is defined by imaginary part of
the dynamical spin susceptibility
 $\quad \chi_{s}^{\prime\prime}({\bf q}, \omega)\quad =  $
  $\quad - \; ({1}/{\pi})\, \mbox{Im}\langle \langle {\bf S}_{q}\mid {\bf
S}_{-q} \rangle \rangle_{\omega + i\delta} \quad $
 which comes from the correlation functions
  $\, \langle B_{1'}(t) B_{2'}(t')\rangle \,$.
\par
For the hole doped case, $\mu \sim \Delta$, the poles of the GF
$F_{\sigma}^{22}({\bf k},\omega_1) $ are close to the Fermi
surface (FS), while  the poles of the GF $ F_{\sigma }^{11}({\bf
k},\omega_1)$ are situated much below the FS at  $\omega_1 \sim
\Delta \gg W$ where $W$ is an effective bandwidth. Therefore, we
can neglect the second term and  use the weak coupling
approximation (WCA) for the first term  in the
self-energy~(\ref{eq:phi-qo}). By taking into account   the
exchange interaction contribution (\ref{eq:x2e}), we get  the
following equation  for the singlet gap $\Phi({\bf q})$:
\begin{equation}
 \Phi({\bf q}) = \frac{1}{N}\sum_{{\bf k}}
  V_{sf}({\bf k, q})
\frac{\Phi({\bf k})}{2 E_{2}(\bf k)} \tanh \frac{ E_{2}(\bf k)}{2
T} ,
 \label{gap22}
\end{equation}
where the interaction $\, V_{sf}({\bf k, q})= J({\bf k-q}) -
 t_{22}^2({\bf k})\, \chi_s({\bf k-q},\omega = 0) $. The
quasiparticle energy in the singlet subband is given by  $\,
E_{2}({\bf k})= [\Omega_{2}^2 ({\bf k})+ \Phi^2({\bf k})]^{1/2}\,$
where $\, \Omega_{2}({\bf k})\,$ is the energy in the normal
state~\cite{Plakida95}.  For an electron doped system $\,(\mu
\simeq 0)\, $ the WCA equation for the gap $\, \Phi^{11}({\bf q})
\,$  is  similar to  Eq.~(\ref{gap22}).

\section{RESULTS AND DISCUSSION}

Let us compare the exchange and spin-fluctuation pairing in the
gap equation~(\ref{gap22}).  The exchange interaction for nearest
neighbors $\,J({\bf q}) = 4 J \gamma({\bf q}),\; \gamma ({\bf q})
= (1/2) (\cos q_x + \cos q_y)\, $.  The spin-fluctuation
contribution is defined by hopping interaction  $t_{22}({\bf k})
\simeq t_{nn}\gamma ({\bf k}) $ with $t_{nn} \sim 0.2$~eV and the
static  susceptibility which we model by the function: $\,
\chi_s({\bf q}) = {\chi_0}/\omega_s (1+\xi^2[1+\gamma({\bf q})])$,
where $\xi$  is the AFM correlation length, $\omega_s \sim J$, and
$\chi_0$ is defined from the normalization condition:
 $\,\langle {\bf S_i}{\bf S_i}\rangle = ({1}/{N})\sum_{\bf q}
\langle {\bf S_{\bf q}}{\bf S_{\bf q}}\rangle =
({3}/{4})(1-\delta),\; \delta = \mid 1-n \mid $.
 \par
 By adopting  a model for the
$d$-wave gap function $\, {\Phi}({\bf q}) = \phi (\cos q\sb{x} -
\cos q\sb{y}) \equiv \phi \eta({\bf q})$ we get from~(\ref{gap22})
the following equation for $\, T\sb{c}$:
\begin{eqnarray}
 1 & = & \frac{1}{N} \sum\sb{\bf k}
 \frac{1} {2  \Omega\sb{2}({\bf k})}
  \tanh \frac{\Omega\sb{2}({\bf k})}{2 T\sb{c}}
\nonumber \\
& \times &\left[ J\,\eta({\bf k})\sp{2}  + \lambda\sb{s}
      \,(4\gamma({\bf k}))\sp{2}
       \eta({\bf k})\sp{2} \right] ,
\label{gap2}
\end{eqnarray}
where $\, \lambda\sb{s}\simeq t_{nn}^2/\omega_s\,$.
 For the exchange interaction mediated by
the interband hopping with a large energy transfer $\Delta \gg W$
the retardation effects are negligible that results in pairing of
all charge carriers in the conduction subband:
\begin{equation}
 T_c \simeq  \sqrt{\mu (W-\mu)} {\exp}{(-1/\lambda_{ex})} ,
 \label{tcex}
\end{equation}
where $\lambda_{ex} \simeq J \, N(\mu)\,$, $N(\mu)$ is the density
of electronic states. For   spin-fluctuation pairing we obtain a
conventional BCS-like formula for $T_c$:
\begin{equation}
T\sb{c} \simeq  \, \omega_s \,
 {\exp}(- {1}/{ \lambda_{sf}})\,  ,
 \label{tcsf}
\end{equation}
where  ${\lambda}_{sf} \simeq \lambda_{s}\, N(\mu)$.   By taking
into account both  contributions, a result similar to (\ref{tcsf})
is obtained,  with an effective coupling constant:
 $\, {\lambda}_{sf}^{(eff)}= \lambda_{sf} +
 {\lambda_{ex}}/[{1- \lambda_{ex} \ln (\mu/ \omega_s)}]\,$.
 By taking for  estimate   $\mu = {W}/2 \simeq 0.35$~eV,
$\, \omega_s \simeq J \simeq 0.13$~eV and $\lambda_{sf} \simeq
\lambda_{ex} =0.2 $ we get $\, \lambda_{sf}^{(eff)} \simeq  0.2 +
0.25 = 0.45\,$ and $\, T\sb{c} \simeq 160$~K, while  the
spin-fluctuation pairing alone gives  $\,T\sb{c}^{0} \simeq
\omega_s \, {\exp}{(-1/\lambda_{sf} })\simeq 10$~K.

A direct  numerical solution of the gap equation~(\ref{gap22})
in~\cite{Plakida03}  proved the $d$-wave symmetry of the gap and
gave the maximum $T_c \sim 0.12 t_{nn}$ for $\,\mu \simeq W/2$ at
an optimal doping $\delta_{opt}\simeq 0.12 $. The spin-fluctuation
interaction produces much lower $T_c$ since it couples the holes
in a narrow energy shell, $\, \omega_s \ll \mu \,$, near the FS.
Moreover,  it vanishes  along  the lines $\, |k_x| +|k_y|=\pi \,$
where $\gamma ({\bf k}) =0$.

 While in electron-phonon superconductors  $T_c$ decreases under
pressure, in cuprates it increases. In particular, in mercury
superconductors $dT_c/da \simeq - 1.35 \times 10^{3}$~K$/$\AA
~\cite{Lokshin01} ($a$ is the in-plane lattice constant) and  $\,
{d \ln T_c}/{d \ln a} \sim - 50 $. From Eq.~(\ref{tcex}) we get an
estimate:
\begin{equation}
\frac{d \ln T_c}{d \ln a}= \frac{d \ln T_c}{d \ln J}\,
 \frac{d \ln
J}{d \ln a} \simeq - \frac{14}{\lambda} \sim - 45,
\end{equation}
for $\lambda_{ex}  \simeq 0.3 $,
 which is quite close to the experimentally observed one.
  Here we take into account that
   $\, J(a) \propto t_{pd}^{4}$ and $\, t_{pd}(a) \propto
1/(a)^{7/2}$.
\par
To explain a small  oxygen isotope effect in cuprates,  $ \alpha_c
= - {d \ln T_c}/{d \ln M} \le 0.1 \,$, we can use
Eq.~(\ref{tcex}). By taking into account the experimentally
observed isotope shift for the N\'{e}el temperature in
La$_2$CuO$_4$~\cite{Zhao94}: $\,
 \alpha_{\rm N} = - {d \ln T_{\rm N} }/{d \ln M} \simeq 0.05 \,
$ with $\,T_{\rm N} \propto J \,$ we get
\begin{equation}
 \alpha_c = - \frac{d \ln T_c}{d \ln M}= - \frac{d \ln T_c}{d \ln
J} \; \frac{d \ln T_{\rm N} }{d \ln M} \simeq 0.16 , \label{tcm}
 \end{equation}
for $\lambda \simeq 0.3 \,$ which is close to experiments.

 To conclude,
the present investigation proves the existence of a singlet
$d\sb{x\sp{2}-y\sp{2}}$-wave superconducting pairing for holes or
electrons in the $p-d$ Hubbard model. It is mediated by the
exchange interaction and antiferromagnetic spin-fluctuation
scattering induced by the kinematic interaction, characteristic to
the Hubbard model. These mechanisms of superconducting pairing are
absent in the fermionic models (for a discussion, see Anderson
\cite{Anderson97}) and are specific  for cuprates.


\begin{thebibliography}{99}

\bibitem{Anderson87}   P.W. Anderson, { Science}, {\bf 235} (1987)
 1196.
\bibitem{Plakida99}
 N.M. Plakida  and  V.S.  Oudovenko,
 { Phys.~Rev.~B} {\bf  59} (1999) 11949.
 \bibitem{Plakida01} N.M.~Plakida, JETP Letters, {\bf  74}
(2001) 36.
\bibitem{Plakida03} N.M. Plakida, L. Anton, S. Adam, and Gh.~Adam,
{ JETP} {\bf 97} (2003) 331.
\bibitem{Plakida95} N.M. Plakida,
R. Hayn,  and J.-L.~Richard,
{Phys.~Rev.~B} {\bf 51} (1995) 16599.
\bibitem{Zubarev60}  D.N. Zubarev,
{ Sov. Phys. Usp.}  {\bf 3} (1960) 320.
\bibitem{Beenen95}  J. Beenen and  D.M. Edwards,
{ Phys. Rev. B} {\bf  52} (1995) 13636.
\bibitem{Avella97} A. Avella,  F. Mancini, D. Villani,  and
H.~Matsumoto,
{Physica C}  {\bf 282-287}  (1997) 1757; T.~Di~Matteo, F.~Mancini,
H.~Matsumoto,  and V.S.~Oudovenko.
{ Physica B,} {\bf  230 - 232} (1997)  915.
\bibitem{Stanescu00}  T.D. Stanescu,   I. Martin, and
Ph.~Phillips,
{ Phys.~Rev.~B} {\bf 62} (2000) 4300.
\bibitem{Lokshin01} K.A. Lokshin, D.A. Pavlov,
S.N.~Putilin,  { et al.}
{ Phys. Rev. B} {\bf  63} (2001) 064511.
 \bibitem{Zhao94}  G.-M. Zhao,  K.K.~Singh, and  D.E.~Morris,
{ Phys. Rev. B} {\bf 50} (1994) 4112.
\bibitem{Anderson97}   P.W. Anderson,
 { Adv. in Phys.} {\bf 46} (1997) 3.
\end{thebibliography}
\end{document}